\documentclass[aps, prb, reprint, longbibliography,superscriptaddress,amsmath,amsfonts]{revtex4-2}

\usepackage{graphicx}
\usepackage{color}
\usepackage{hyperref}
\usepackage{xcolor}
\usepackage{comment}
\usepackage[normalem]{ulem}
\usepackage{enumitem,amssymb}
\usepackage{times}
\usepackage{pifont}

\newlist{todolist}{itemize}{2}
\setlist[todolist]{label=$\square$}

\def\ket#1{| #1 \ra }
\def\bra#1{\la #1}

\def\la{\langle}
\def\ra{\rangle}

\def\beq{\begin{equation}}
\def\eeq{\end{equation}}
\def\bea{\begin{eqnarray}}
\def\eea{\end{eqnarray}}
\def\barr{\begin{array}}
\def\earr{\end{array}}

\def\Id{\mathbb{I}}

\definecolor{MyDarkBlue}{rgb}{0,0.08,0.45} 
\definecolor{MyLightMagenta}{cmyk}{0.1,0.8,0,0.1} 
\definecolor{MLM}{cmyk}{0.1,0.8,0,0.1} 
\definecolor{MyDarkGreen}{rgb}{0,0.45,0.08} 
\definecolor{MDG}{rgb}{0,0.55,0.05}

\begin{document}

\title{Dynamical quantum phase transition and thermal equilibrium in the lattice Thirring model} 

\author{Mari~Carmen~Ba\~{n}uls}
\email{banulsm@mpq.mpg.de}
\affiliation{Max-Planck Institut f\"{u}r Quantenoptik, Garching 85748, Germany}
\affiliation{Munich Centre for Quantum Science and Technology (MCQST), Schellingstrasse 4, Munich 80799, Germany}

\author{Krzysztof~Cichy}
\email{krzysztof.cichy@gmail.com}
\affiliation{Faculty of Physics and Astronomy, Adam Mickiewicz University, Uniwersytetu Pozna\'{n}skiego 2, 61-614 Pozna\'{n}, Poland}

\author{Hao-Ti~Hung}
\email{hunghaoti852@gmail.com}
\affiliation{Department of Physics, National Taiwan University, Taipei 10617, Taiwan}
\affiliation{Center for Theoretical Physics, National Taiwan University, Taipei 10617, Taiwan }

\author{Ying-Jer~Kao}
\email{yjkao@phys.ntu.edu.tw}
\affiliation{Department of Physics, National Taiwan University, Taipei 10617, Taiwan}
\affiliation{Center for Theoretical Physics, National Taiwan University, Taipei 10617, Taiwan }
\affiliation{Center for Quantum Science and Technology, National Taiwan University, Taipei 10617, Taiwan  }

\author{C.-J.~David~Lin}
\email{dlin@nycu.edu.tw}
\affiliation{Institute of Physics, National Yang Ming Chiao Tung University, Hsinchu 30010, Taiwan}
\affiliation{Centre for High-Energy Physics, Chung-Yuan Christian University, Chung-Li 32023, Taiwan}

\author{Amit~Singh}
\email{amitletit@gmail.com}
\affiliation{Department of Mechanical Engineering, National Yang Ming Chiao Tung University, Hsinchu 30010, Taiwan}
\affiliation{Department of Physics and Astronomy, University of Manchester, Manchester M13 9PL, United Kingdom}

\begin{abstract}
Using tensor network methods, we simulate the real-time evolution of the
lattice Thirring model quenched out of equilibrium in both the critical and
massive phases and study the appearance of dynamical quantum phase
transitions, as nonanalyticities in the Loschmidt rate.
Although the presence of a dynamical quantum phase transition in the model does
not correspond to quenches across the critical line of the equilibrium phase
diagram at zero temperature, we identify a threshold in the energy density of
the initial state, necessary for a dynamical quantum phase transition to be
present. Moreover, in the case of the gapped quench Hamiltonian, we unveil a
connection of this threshold to a transition between different regions in the
finite-temperature phase diagram.
\end{abstract}
\maketitle

\section{Introduction}
\label{sec:intro}
Out-of-equilibrium dynamics is one of the most challenging problems in the
study of quantum many-body systems. This funnels the interest towards finding
universal behaviors that allow a more comprehensive understanding of the
dynamics. A paradigmatic scenario to investigate these questions is that of a
quantum quench~\cite{AditiMitra_2018}, in which a system is evolved with a
Hamiltonian of interest, after having been initialized in a certain (often
pure) state. Since the mid-2010s, an active direction of research in this
context has been the investigation of dynamical quantum phase transitions 
(DQPTs)~\cite{Heyl2012dqpt,Heyl2018review}, which appear as zeros in the 
Loschmidt echo or return probability, i.e., the probability that the system is 
found in the initial state during the evolution.

One of the problems for studying out-of-equilibrium dynamics of such complex
systems is the scarcity of appropriate tools. Only in certain cases is an
analytical solution possible that allows a complete description of the time
evolution after a quench~\cite{Heyl2012dqpt,Torlai_2014,Uhrich_2020}.
For the most general cases, however, the only possibility is to obtain an
approximation to the dynamics by means of numerical simulations, within the
regimes where these are feasible. A relevant role among numerical methods is
played by tensor network state (TNS) techniques~\cite{Verstraete2008,
Schollwoeck2011,Orus2014annphys,Silvi2019tn,Okunishi2022,Banuls2023}.
In particular, in one-dimensional systems, matrix product state
(MPS)~\cite{fannes92fcs,Oestlund1995,Vidal2003,Verstraete2004,
Perez-Garcia2007} algorithms provide an effective way to explore the
time-dependent properties of the Loschmidt echo for systems in the
thermodynamic limit~\cite{Andraschko2014,Heyl2018review}. Although the
potentially fast entanglement growth~\cite{Calabrese:2004eu,Osborne2006,
Schuch2008a}limits the time that can be reliably simulated,
TNS dynamical methods~\cite{Haegeman2016unif,Vanderstraeten2016,
Paeckel2019tevol} enable an accurate picture of the quantity of interest for a
moderate time after the quench, which has been exploited to obtain important 
results regarding DQPTs~\cite{Andraschko2014,Karrasch_2017,Zauner2017,
DeNicola2021ent,Halimeh_2021,Halimeh2015,VanDamme_2023,
osborne2024mesonmasssetsonset}.

Signatures of DQPT have been observed experimentally in trapped ion quantum
simulators~\cite{Jurcevic_2017} and superconducting qubits~\cite{Guo_2019}.
While initial studies suggested that DQPTs occur when the parameter of the
Hamiltonian is quenched across an equilibrium phase boundary~\cite{
Heyl2012dqpt, Heyl2018review, Karrasch_2017}, it was later realized that
this is not the general case~\cite{Vajna_2014,Andraschko2014,Halimeh2017,
Homrighausen2017,Corps2022}, and the nature of DQPTs is more complex and
far from being completely understood~\cite{VanDamme_2023,PerezGarcia2024dqpt}.

The study of out-of-equilibrium physics for quantum field theories ({QFTs}),
defined in the continuum, is highly nontrivial. Equilibrium properties of
such models are usually addressed by the path integral formalism and Monte
Carlo sampling, which, however, encounters a sign problem in certain scenarios.
This limitation has motivated an investigation of applicability of TNS methods
to the realm of QFT. In particular, gauge theories in 1+1 dimensions have been
systematically addressed using TNS methods (see,
e.g., Refs.~\cite{Banuls2020ropp,Banuls2020qtflag,bass2021qtech} and references
therein). Other QFTs and conformal field theories have also been studied
in the same manner~\cite{Banuls:2019hzc,Hauru_2016,Hu_2017,Vanhove_2018}. These
investigations have focused on the equilibrium properties of the field
theories. Real-time simulations are more scarce~\cite{Buyens:2013yza,
kuehn2015string,Buyens:2015tea,Pichler:2015yqa,schmoll_2023}, and studying the real-time
dynamics of a continuum QFT with these methods remains an open problem.

\begin{figure}[t]
  \centering
  \includegraphics[width=\columnwidth]{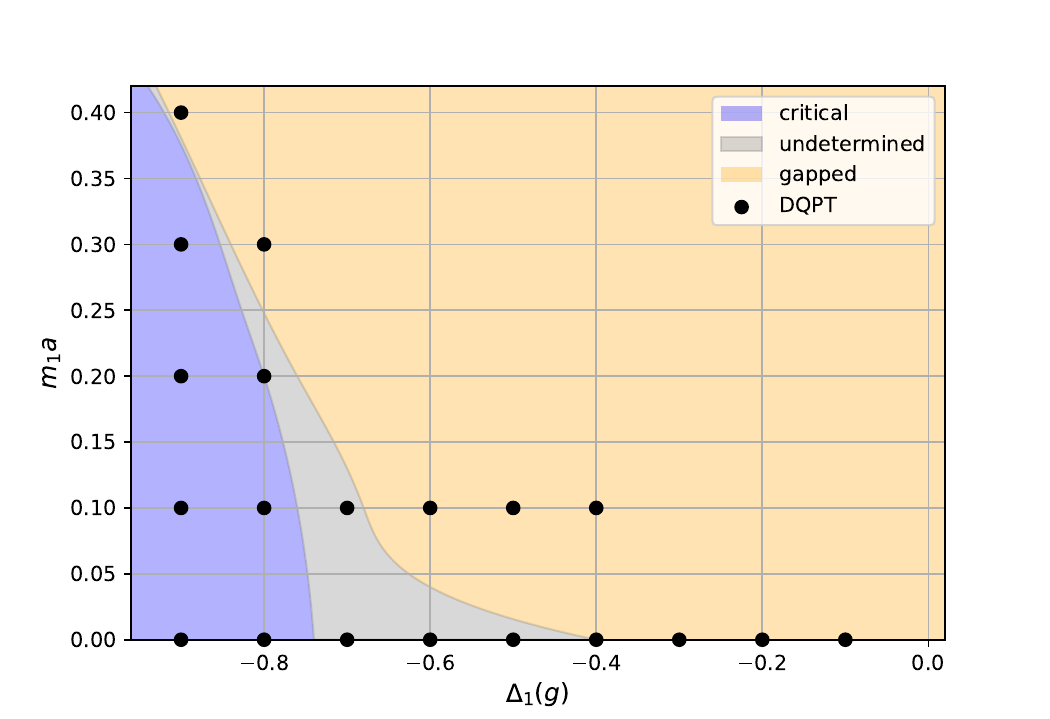}
  \caption{Zero-temperature phase diagram of the Thirring model in the
  coupling-mass plane. The color-coded areas on the plot depict the equilibrium
  phase through correlator analyses~\cite{Banuls:2019hzc}. The blue region is the
  critical region, where the constant $C$ [defined in Eq.~\eqref{eq:corr_ansatz}]
  is less than $0.001$; the orange region is the gapped phase region, where $C>0.01$;
  the gray region is undetermined, where the corresponding $C\in[0.001, 0.01]$.
  The black point marks the location where we observe the DQPT, starting from the
  ground state of the Hamiltonian $H_1(m_1a, \Delta_1)$ and quenching to
  $H_2(m_2a=0.1, \Delta_2=0.9)$.}
  \label{fig:equ_phase_dqpt}
\end{figure}

The most straightforward way to study QFT problems with TNS is to use the
discretized version of the model Hamiltonian and apply standard TNS
methods to find their equilibrium or dynamical properties. Such results
depend on different  lattice discretization schemes, but by repeating
the procedure at different lattice spacings and performing the proper
renormalization, it is possible to recover the continuum limit 
(see, e.g., Refs. ~\cite{Banuls2013a,Buyens:2013yza}). In this work, we focus on one
such  spin chain, corresponding to the discretization of the 1+1-
dimensional Thirring model. The resulting spin model is a specialized XXZ
Hamiltonian coupled to both staggered and homogeneous magnetic fields. The
ground state of the model, as determined in previous work~\cite{Banuls:2019hzc}
using MPS, exhibits a critical and a gapped phase, separated by a
Berezinskii-Kosterlitz-Thoules (BKT) phase transition
(see Fig.~\ref{fig:equ_phase_dqpt}). For vanishing bare fermion mass, the model
is critical for all couplings.

In this work, we use uniform MPS to study the phenomenology of DQPTs in the
spin model for the discretized Thirring model. We simulate quenches in both
phases from initial states with very different properties. We show that a
necessary condition for a DQPT to occur within the finite-time window that
our numerical method can access is that the energy density of the initial
state is above a certain threshold, independent of the phases of the initial
and quenched Hamiltonian. However, in the case of the quench to the gapped
phase we also identify a connection of DQPTs to the equilibrium phase
diagram but at finite temperature. Namely, at the temperature corresponding
to the identified energy density threshold, the properties of the thermal
equilibrium state undergo a substantial change, with the vanishing of a
string correlator that is nonzero in the ground state. We further show that
the times at which DQPTs occur as a function of the energy density of the
initial state show a structure reminiscent of the complex zeros of the
Loschmidt amplitude~\cite{Peotta2021cumulant,McCulloch2023dqpt} and several
branches of DQPTs that can be explored from different initial states.

The rest of the paper is structured as follows. In Sec.~\ref{sec:formalism}
we introduce the spin model that corresponds to the discretized Thirring
Hamiltonian and present the MPS formulation and the tools and algorithms
we have used. Second, we review the equilibrium phase structure of the
Thirring model from the previous research in Sec.~\ref{sec:equil_phase_struct}.
In Sec.~\ref{sec:loschmidt}, we discuss the DQPT concept and the use of MPS
methods to study them in the model of interest. Our results for DQPTs in
different quenches, and the relation of their appearance with the finite
temperature phase diagram are discussed in Sec.~\ref{sec:temperature}. Finally,
in Sec.~\ref{sec:conclusion} we summarize our findings and discuss potential
further investigations.

\section{Formalism}
\label{sec:formalism}
\subsection{The Thirring model as a spin model}
\label{sec:thirring_spin}
In this work we focus on the spin chain formulation corresponding to the lattice 
discretization of the massive Thirring model in 1+1 dimensions.
The action of the original field theory
is described by
\begin{align}
S_{\text{Th}}\left[\psi, \bar{\psi}\right] = &\int d^2x \Big[ 
 \bar{\psi} i \gamma^\mu \partial_\mu \psi 
- m \bar{\psi} \psi \nonumber \\
& - \frac{g}{2} (\bar{\psi} \gamma_\mu \psi)(\bar{\psi} \gamma^\mu \psi)
\Big],
\label{eq:Thirring_model_action}
\end{align}%
where $m$ is the fermion mass and $g$ is the four-fermion coupling.
To facilitate the use of MPS, quantization of the classical theory
described by Eq.~(\ref{eq:Thirring_model_action}) is performed through the
canonical method in the Hamiltonian formalism. One subtlety in this formalism
is the inclusion of effects of the anomaly in the four-fermion operator.
These effects can be easily investigated through the examination of the
path-integral measure in the Lagrangian formulation~\cite{Das:1985ci}.
In the canonical procedure, they need to be accounted for via a nonlocal
definition of the currents in the four-fermion operator~\cite{Schwinger:1962tp,
hagen1967new}. 

The model can be discretized on a one-dimensional {spatial} lattice
using the staggered regularization. For convenience of carrying out numerical
computations in this work, the fermionic degrees of freedom in this
Hamiltonian are mapped onto spin operators through the Jordan-Wigner 
transformation. Details of the above procedure can be found in 
Ref.~\cite{Banuls:2019hzc}. Implementation of this strategy turns the
Hamiltonian operator of the continuum Thirring model, $H_{{\mathrm{Th}}}$,
into that of the XXZ spin chain coupled to both uniform and staggered magnetic
fields, up to a scaling factor,
\begin{align}
\label{eq:H_sim}
H(\tilde{m}_{0}a, \Delta) =& -\frac{1}{2} \sum_{n}^{N-2}\left( S_{n}^{+}S_{n+1}^{-} 
            + S_{n+1}^{+}S_{n}^{-} \right)\nonumber \\
            +& a \tilde{m}_{0} \sum_{n}^{N-1} \left(-1\right)^{n} \left(
              S_{n}^{z}+\frac{1}{2} \right) \nonumber  \\
        +&\Delta (g) \sum_{n}^{N-1} \left( S_{n}^{z}+\frac{1}{2} \right) \ 
            \left( S_{n+1}^{z}+\frac{1}{2} \right) \,,
\end{align}%
where $a$ is the lattice spacing, $N$ is the total number of lattice
sites, $S_{n}^{\pm} = S_{n}^{x} \pm iS_{n}^{y}$ and $S_{n}^{z}$ are the spin
matrices ($S^{i}_{n} = \sigma^{i}/2$ with $\sigma^{i}$ being the Pauli 
matrices) at the $n$th site, and $[S^{i}_{n}, S^{j}_{m}]_{n\not=m} = 0$.  
Here $\tilde{m}_{0} = m_{0}/\nu (g)$ with $m_{0}$ being the bare counterpart
of the mass parameter, $m$, in Eq.~(\ref{eq:Thirring_model_action}). The
functions $\nu (g)$ and $\Delta (g)$ are the lattice version of wave-function
renormalization and the four-fermion coupling~\cite{Luther:1976mt}, 
\begin{align}
\label{eq:nu_and_Delta}
 \nu (g) =& \left ( \frac{\pi - g}{\pi} \right )/ \sin\left (
 \frac{\pi - g}{2} \right ) \, , \nonumber\\
   &\Delta (g) = \cos \left (
 \frac{\pi - g}{2} \right )  \, .
\end{align}%
Notice that the physical Hamiltonian is obtained by rescaling the
dimensionless one in Eq.~\eqref{eq:H_sim} by a factor $\nu(g)/a$.
Throughout the rest of the paper,  we will work with the lattice
Hamiltonian ~\eqref{eq:H_sim} and correspondingly define dimensionless
time and inverse-temperature parameters, which involve rescaling the
physical (dimensionful) magnitudes by $\nu(g)/a$.  Additionally, we
restrict our consideration to the sector with total spin (magnetization)
zero~\cite{Banuls:2019hzc} and find the ground state by adding the
penalty term (see Appendix ~\ref{appdx:penalty_mpo}). 

\subsection{Matrix product state}
\label{sec:mps}
A generic quantum state for a system of $N$ spins can be written in the form
\beq
\label{eq:N_spin_state}
 |\Psi_{N} \rangle = \sum_{\sigma_{1},\cdots,\sigma_{N}} 
 c_{\sigma_{1}\cdots\sigma_{N}}\,|\sigma_{1}\cdots\sigma_{N}\rangle \,,  
\eeq
which can then be approximated as a MPS with bond dimension $D$,
\beq
\label{eq:finite_MPS_N_spin_state}
   |\Psi_{N} \rangle = \sum_{\sigma_{1},\cdots,\sigma_{N}} \mathrm{tr}
   \left[M_1^{\sigma_{1}}M_2^{\sigma_{2}} \cdots 
   M_N^{\sigma_{N}}\right]\,|\sigma_{1}\cdots\sigma_{N}\rangle \,,
\eeq
where $M_i^{\sigma_{i}}$ is a $D\times D$ matrix. 
For the open boundary condition,  $M_1^{\sigma_{1}}$ and $M_N^{\sigma_{N}}$
become $D-$dimensional vectors. In this case, the trace in 
Eq.~(\ref{eq:finite_MPS_N_spin_state}) becomes a simple product of matrices.  

We search for DQPTs in the real-time evolution of Eq.~\eqref{eq:H_sim}.  
For this purpose, it is desirable to perform calculations directly in the
thermodynamic limit. This can be accomplished by employing the uniform MPS
(uMPS) technique, which allows for the description of the state of an
infinite spin chain with translation invariance using one bulk tensor and
two appropriate boundary tensors~\cite{Phien2012, Zauner2018vumps}. The
uniform MPS can be represented as 
\beq
 \label{eq:uMPS}
   |\Psi(A)\ra = \sum_{\pmb{\sigma}} {\left (\cdots 
   A^{\sigma_{n-1}}A^{\sigma_{n}}A^{\sigma_{n+1}}A^{\sigma_{n+2}}
   \cdots\right)}|\pmb{\sigma}\ra \, ,
\eeq
where $A^{\sigma_{n}}\in \mathbb{C}^{D\times D}$ for $\sigma=1,\cdots, d$.
It can be represented diagrammatically as
\begin{equation}
\label{eq:uMPS_unit}
 \includegraphics[scale=0.9]{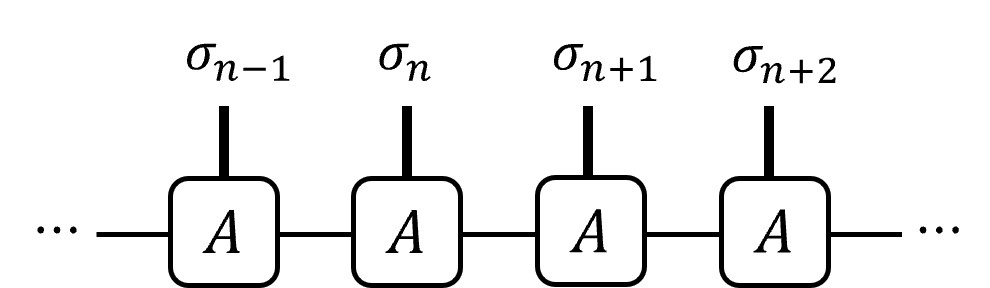} ,
\end{equation}
where $A$ is the unit cell of the uMPS.
Due to the staggered term in  Eq.~\eqref{eq:H_sim}, 
the system is translationally invariant with a period of two. Thus, we choose
the tensor $A$ to represent two contiguous sites of the chain, and its 
physical dimension becomes $d=4$.
\section{Equilibrium phase structure of the 1+1-dimensional Thirring model }
\label{sec:equil_phase_struct}
In this section, we review our study of the zero-temperature phase structure
of the Thirring model using MPS \cite{Banuls:2019hzc}. An analysis of
renormalization group (RG) equations of the dual sine-Gordon
theory~\cite{Amit:1980,Kaplan:2009kr,ZinnJustin:2000dr} points to the existence
of two phases in the Thirring model. The massless theory is a conformal field
theory at all couplings $\Delta(g)$. RG flows in the mass-coupling plane reveal
that the line $m=0$ is a fixed line under RG transformations. However, it is 
stable/unstable for couplings $\Delta(g)$ below/above certain
$\Delta(g_\star)_{m=0}=-\pi/2$. At $m>0$, the nonzero fermion mass is an
irrelevant operator for $\Delta(g)<\Delta(g_\star)_m$ but becomes relevant
when $\Delta(g)>\Delta(g_\star)_m$, with the transition coupling
$\Delta(g_\star)_m$ being mass dependent. This gives rise to two equilibrium
phases:\\
\hspace*{15 pt}(i) \quad $\Delta(g)<\Delta(g_\star)_m$ : gapless (critical) phase\\
\hspace*{15 pt}(ii) \quad  $\Delta(g)>\Delta(g_\star)_m$ : gapped (massive) phase,\\
separated by a BKT transition. The BKT transition was originally discussed in
the context of the XY model \cite{Kosterlitz:1973xp,Kosterlitz:1974sm}, which
is actually dual to the sine-Gordon and Thirring theories, with the XY model
temperature dual to the couplings of these theories. For a more extensive
discussion of these aspects, including von Neumann entanglement entropy and
chiral condensate, we refer to Ref.~\cite{Banuls:2019hzc}.

The quantitatively most robust analysis of the phase structure and the
determination of $\Delta(g_\star)_m$ is about the string correlator,
\begin{align}
\label{eq:Cstring}
 C_{\rm string}(x) = &\frac{1}{N_x}\sum_n \langle S^z(n)S^z(n+1) \cdots \nonumber\\
    &\cdots S^z(n+x-1) S^z(n+x) \rangle,
\end{align}
which is a string of $S^z$ operators. We note that Ref.~\cite{Banuls:2019hzc}
erroneously states this string correlator as the fermion-antifermion one,
containing $\langle S^+(n)S^z(n+1) \ldots S^z(n+x-1) S^-(n+x) \rangle$ terms.
At large-enough distances, the correlator $C_{\rm string}(x)$ exhibits a
power-law decay in the gapless phase, given by
$C_{\rm string}(x)\propto\beta x^{-\alpha}$, and a power-exponential decay in
the gapped phase, $C_{\rm string}(x)\propto B x^{-\eta} A^x$. The crucial
distinction that makes the string correlator particularly useful for
determining $\Delta(g_\star)_m$ is that it decays to a constant, $C$, in the
gapped phase. In practice, one can analyze both phases based on the
power-exponential fitting ansatz:
\begin{equation}
C_{\rm string}(x) = B x^{-\eta} A^x + C,
\label{eq:corr_ansatz}
\end{equation}
with $A=1$ found in fits in the gapless phase.
Thus, the phase transition when going from the gapless into the gapped phase is
manifested by $A$ dropping below 1 and, in the case of the string correlator,
by $C>0$. The onset of the nonzero constant is numerically more robust than 
the smooth drop of $A$ below 1, accounting for the practical usefulness of the
string correlator. As we will see below, this is true also at nonzero
temperatures and it will allow us to identify relations between thermal phases
of the Thirring model and DQPTs.

As a summary of our findings for the zero-temperature phase structure, we show
Fig.~\ref{fig:equ_phase_dqpt}. The color-coded regions correspond to the
equilibrium phase on the coupling-mass plane through the correlator analyses
from our previous work \cite{Banuls:2019hzc}. In particular, we fitted
Eq.~(\ref{eq:corr_ansatz}) and extracted values of the constant $C$. Since the
results depend, in general, on the fitting interval, we adopted a systematic
procedure to assess the systematic uncertainty of $C$ (dominating with respect
to other uncertainties, coming from the finite bond dimension and the finite
volume). For details of this procedure, we refer again to Ref.~\cite{Banuls:2019hzc}.
Depending on the value of $C$, we identified three situations. With $C<0.001$
or $C>0.01$, one unambiguously concludes either the critical phase (purple
region) or the gapped phase (orange region), respectively. If $C\in[0.001,0.01]$
(gray region), then the systematic uncertainty does not allow us to draw conclusions
about a (non)vanishing value of $C$. This region of ambiguous values of $C$ is
indicated by gray shading and the BKT transition must occur there. We note the
onset a nonzero value of $C$ is very smooth at small fermion masses, while it
becomes rather sharp with an increasing mass. At very small masses, the implied
transition point is consistent with the RG flow analysis [$\Delta(g)\approx-\pi/2$]
and at large masses, it moves towards more negative values.

\section{The Loschmidt echo and the mixed transfer matrix}
\label{sec:loschmidt}
\subsection{Quantum quench and DQPT}
In this work, we investigate DQPT in a quantum quench. A quantum quench is the
paradigmatic scenario where to study nonequilibrium dynamics of a quantum
system. The system is assumed to be prepared in a certain initial state
$\ket{\Psi_1}$, typically the ground state of a Hamiltonian $H_1$. From time
$t=0$ on, it is evolved with a different Hamiltonian $H_2$. The evolved state
can be written as $\ket{\Psi(t)}=e^{-itH_2}\ket{\Psi_1}$. The key ingredient in
the study of DQPT is the Loschmidt amplitude,
\beq
\label{eq:loschimidt_amp_def}
 G(t)= \bra{\Psi_{1}}\ket{\Psi(t)}= \la \Psi_{1} | e^{-i H_2 t} | \Psi_{1}\ra \, ,
\eeq
which is analogous to the boundary partition function~\cite{LeClair1995}
$Z_B(z)=\langle\Psi_{1}|e^{-zH_2}|\Psi_{1}\rangle$ in the complex plane
$z\in\mathbb{C}$. The nonanalytical points of the free energy density
$f(z)=-\lim_{N\to\infty}\frac{1}{N} \log Z_B(z)$ are believed to indicate phase
transitions~\cite{Saarloos_1984}. The squared modulus of Loschmidt amplitude in
Eq.~\eqref{eq:loschimidt_amp_def}, the Loschmidt echo, quantifies the return
probability, i.e., the probability of the system being found in the initial
state after a certain time $t$. In the thermodynamic limit
($N\rightarrow \infty$), the Loschmidt echo will approach $e^{-N r(t)}$, with
the rate function
\beq
r(t)=-\lim_{N\to\infty}\frac{1}{N} \log |G(t)|^2.
\label{eq:Loschmidt_rate}
\eeq
This function can be interpreted as a dynamical free energy, and can exhibit
nonanalyticities. DQPTs are defined as
the times when $r(t)$ is nonanalytical~\cite{Heyl2012dqpt}. In some cases,
DQPTs are associated with quenches across an equilibrium phase
transition~\cite{Heyl2012dqpt,Karrasch2013nonint}, but such connection to
equilibrium is not always present~\cite{Vajna_2014}, which motivates the
identification of DQPTs as a genuinely nonequilibrium effect. The potential
observable signature of DQPTs in local and string quantities has also been a
subject of intensive studies.~\cite{Bandyopadhyay_2021, Bandyopadhyay_2023}
The complex phenomenology of DQPTs  has motivated studiees in multiple models, 
but the phenomenon is yet far from a full theoretical
understanding~\cite{Heyl2018review,Andraschko2014,Halimeh_2021,Bhat_2024,Zhao_2024}.

\begin{figure*}[ht]
  \includegraphics[width=\linewidth, trim = 5 -35 0 0]
  {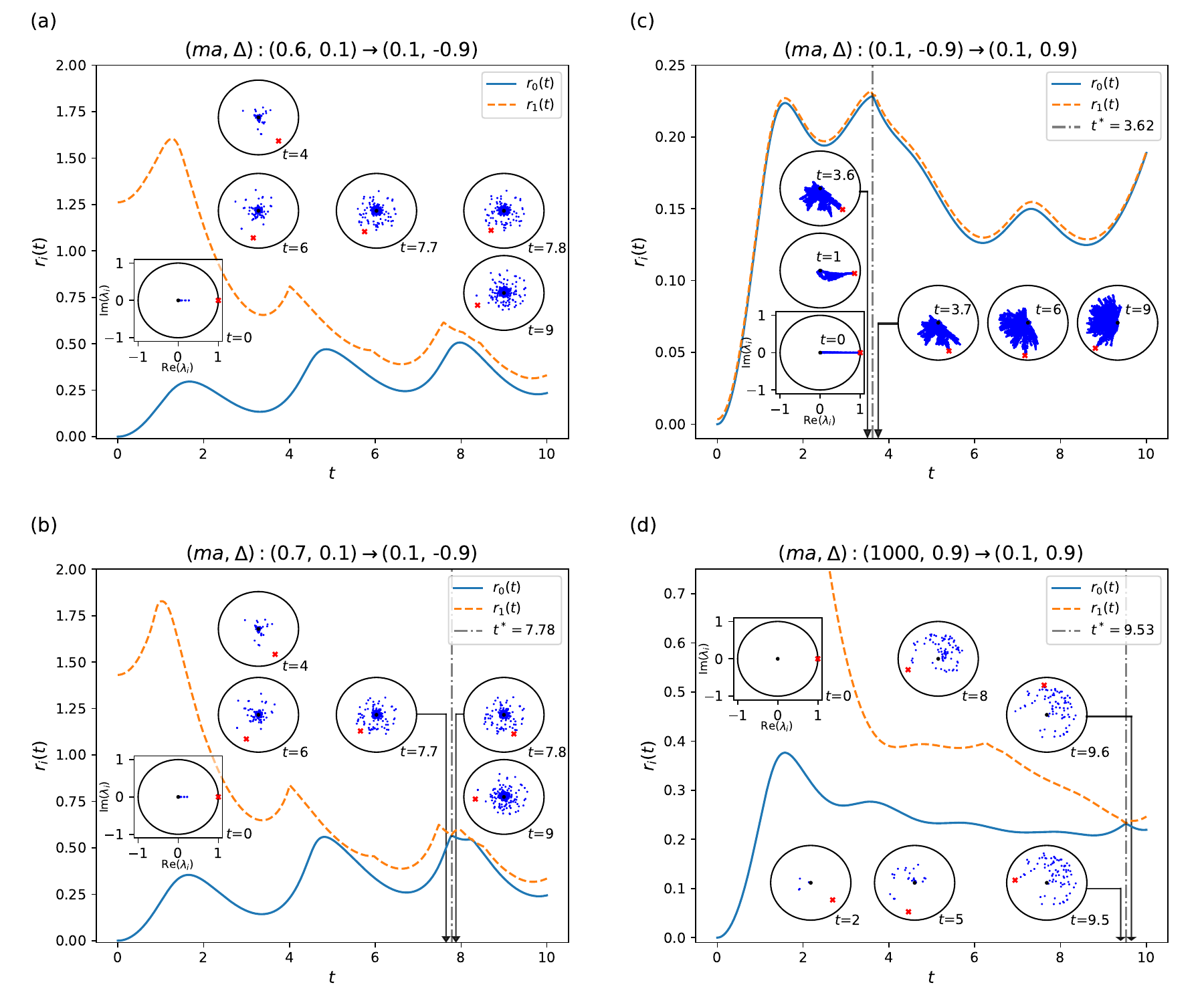}
  \caption{The return rate function and the full spectrum of the mixed transfer
  matrix $T(t)$. The quantity $r_i$ is defined in Eq.~\eqref{eq:return_spctrum},
  while $r_0$ corresponds to  the return rate function defined in
  Eq.~\eqref{eq:Loschmidt_rate}. The insets display the full spectrum of the
  mixed transfer matrix $T(t)$  defined in Eq.~\eqref{eq:mixed_transfer_op_def},
  with the red cross representing the corresponding dominant eigenvalue. The
  time $t^{*}$ (indicated by the gray dash-dotted line) denotes when the first
  DQPT occurs. This plot shows how the nonanalyticities in the return rate
  $r_0(t)$ correspond to discontinuous jumps in the phase of the dominant
  eigenvalue of $T(t)$ (red point). We can notice that in (a), with
  $(m_1 a, \Delta_1)=(0.6, 0.1)$ and $(m_2 a, \Delta_2)=(0.1, -0.9)$, there are
  no DQPTs. However, in (b), a small increase in $m_1 a$ to $0.7$, with the same
  quench Hamiltonian, induces a DQPT at $t^{*}=7.78.$}
  \label{fig:selected_quenches}
\end{figure*}

The transfer matrix formalism allows for a systematic exploration of the
character of DQPTs~\cite{Andraschko2014,Zauner2017,DeNicola2021ent,Halimeh_2021,
VanDamme_2023}. We focus here on the case where the initial state is a MPS 
and consider directly the translationally invariant case, i.e., the initial
state is parametrized by a single rank-3 tensor $A$. The Loschmidt amplitude,
for a translationally invariant system of $N$ sites, can then be expressed as
\beq
 G(t)=(l| T^{N}(t) | r)={\lambda_0^N},
 \label{eq:LoschG}
 \eeq
where $T(t)$
is the {mixed transfer matrix} defined by
\beq
T(t):=\sum_i {\bar{A}(0)^{\sigma}}\otimes A(t)^{\sigma},
\label{eq:T(t)}
\eeq
 and $A(t)$ represents the tensor describing the evolved state. The vectors
 $(l|$ and $|r)$ are the left and right eigenvectors with the leading
 eigenvalue {$\lambda_0$}, satisfying $(l|r)=1$. Graphically,
\begin{equation}
    \label{eq:mixed_transfer_op_def}
    \includegraphics[scale=0.88]{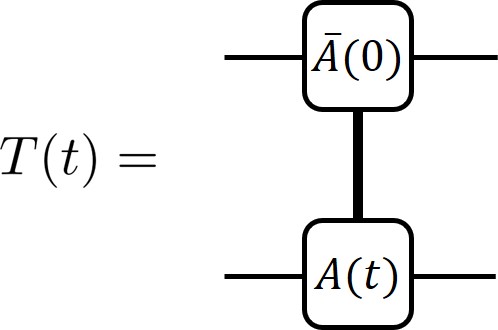}.
\end{equation}
The Loschmidt rate is thus determined by the spectrum of the (in general
non-Hermitian) matrix $T(t)$ and, in the thermodynamic limit, by its dominant
eigenvalue {$\lambda_0(t)$}. DQPTs appear when there is an eigenvalue
crossing and the eigenvector associated to the dominant eigenvalue
changes~\cite{Andraschko2014, Hamazaki2021}.

\subsection{Investigation of DQPT in the lattice version of the Thirring model}

\begin{figure*}[t]
\centering
  \includegraphics[width=\linewidth]{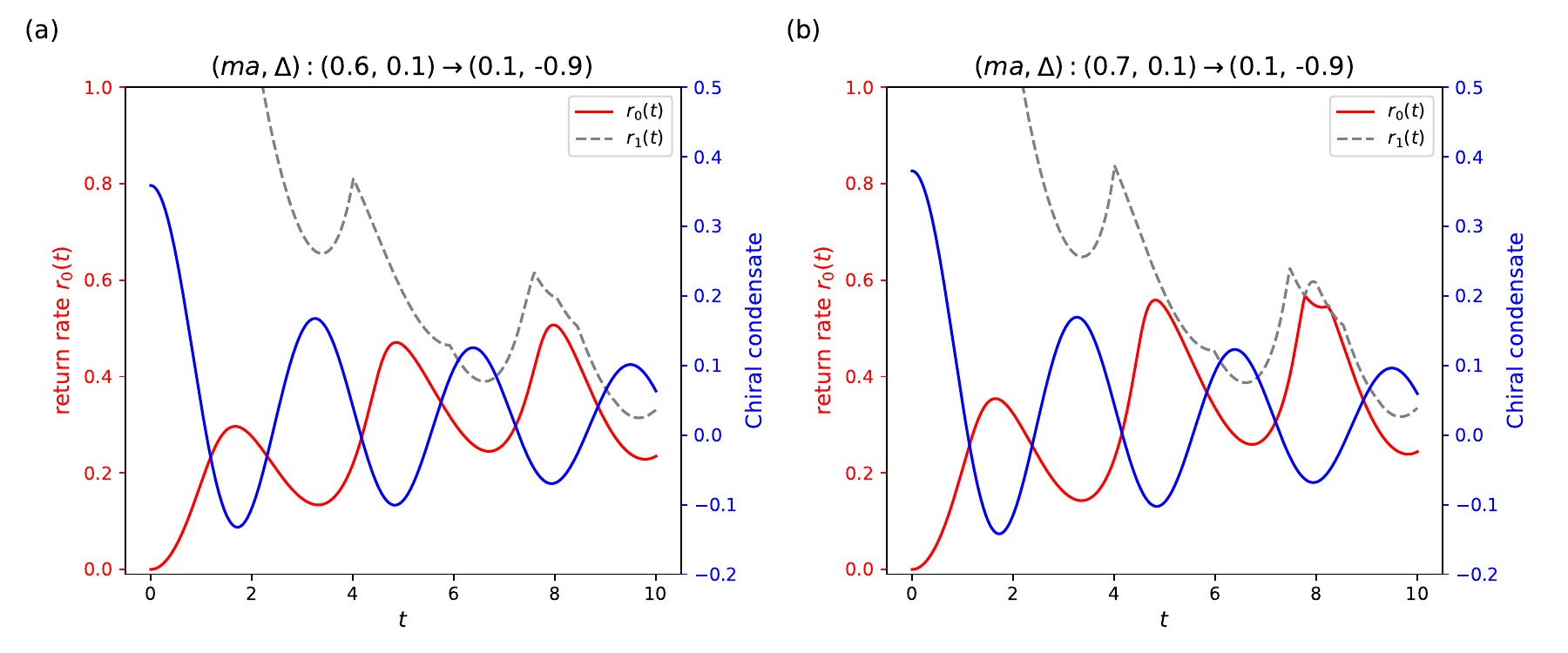}
  \caption{The return rate functions (red)  and the chiral condensate (blue) for 
  the cases  (a)  without and (b) with DQPTs. The parameters we choose here are 
  the same as those in Figs.~\ref{fig:selected_quenches}(a) and 
  \ref{fig:selected_quenches}(b).}
  \label{fig:reply1}
\end{figure*}
Here we focus on the Hamiltonian {Eq.~\eqref{eq:H_sim}}
and consider quenches from the ground state  $\ket{\Psi_{1}}$ of the model with
a given set of parameters, $H_1=H(m_1 a,\Delta_1)$. The quench Hamiltonian,
with which the state is evolved, will be denoted as $H_2=H(m_2 a,\Delta_2)$.
For a fixed $H_2$, we study the real-time evolution of different initial states,
$\ket{\Psi_{1}}$. The latter are represented by their MPS approximation,
obtained using the variational uMPS (VUMPS) algorithm~\cite{Zauner2018vumps},
and the real-time evolution according to $H_2$ can be numerically simulated
using a standard tensor network algorithm~\cite{Schollwoeck2011,Orus2014,
Banuls2023}. In particular, we use time-dependent variational principle (TDVP)
algorithm~\cite{Haegeman2011itdvp,Haegeman2016unif} to obtain the time-dependent
tensors of the MPS representation, $A(t)$. We obtain the spectrum of the mixed
transfer matrix, $T(t)$, using Eq.~\eqref{eq:T(t)} and explore the appearance
of DQPTs for various quench parameters. To investigate how DQPTs occur, we
introduce the following quantity:
\beq
\label{eq:return_spctrum}
r_i(t)=-\log{|\lambda_i(t)|^2},
\eeq
where $\{\lambda_i\}$ is the spectrum of the mixed transfer matrix $T(t)$,
satisfying $|\lambda_0|\geq|\lambda_1|\geq\cdots$. Notice that $r_0$ is exactly
return rate function defined in Eq.~\eqref{eq:Loschmidt_rate}.
Figure~\ref{fig:selected_quenches} shows the results for a selection of
simulation parameters. The first DQPT occurs when $t=t^{*}$. The insets show
snapshots of the spectrum of $T(t)$, which demonstrate that the
nonanalytical points in the return rate function in 
Eq.~\eqref{eq:Loschmidt_rate} are associated with a discontinuous jump in the
phase of the dominant eigenvalue of {this} mixed transfer 
matrix~\cite{Hamazaki2021}.

By choosing parameters, $(m_1 a, \Delta_1)$ and $(m_2 a,\Delta_2)$, to be in a
different or in the same equilibrium phase, we can simulate quenches across the
phase diagram, which we determined in our previous study~\cite{Banuls:2019hzc}.
We find that, as observed in other models~\cite{Vajna_2014,Andraschko2014},
the equilibrium {phases} of $H_1$ and $H_2$ does not determine the presence of
DQPTs, which can appear in both types of scenario. In
Fig.~\ref{fig:equ_phase_dqpt}, we illustrate that the existence of DQPTs is not
equivalent to the equilibrium phase. DQPTs have been demonstrated to occur in
quenches that happen entirely within either of the equilibrium phases.

\subsection{DQPT and thermalization}

Notice that DQPTs and (pre-)thermalization happen at different timescales. 
In our setup, DQPTs can occur at relatively short times. Hence, we do not expect 
that local observables have equilibrated yet.
Indeed, we observe that local observables are still varying strongly at the time 
slices when the DQPTs are observed. To illustrate this, we plot in 
Fig.~\ref{fig:reply1} the time evolution of the chiral condensate (or the 
staggered magnetization in spin model language), 
$$
{\chi}\sim  \frac{1}{N}\left|\sum_n{(-1)^n\left( 
S_n^z +\frac{1}{2} \right)}\right|,
$$
 in both the cases with or without DQPT.
 The chiral condensate shows strong time variation at the longest time allowed 
 for our simulation, $t=10$, in both cases, indicating DQPT and prethermalization 
 or thermalization phenomena  occur at different timescales.

\section{DQPTs and finite-temperature phase diagram}
\label{sec:temperature}
\subsection{Energy threshold and the inverse temperature}

\begin{figure*}[ht]
\centering
\includegraphics[width=\linewidth]{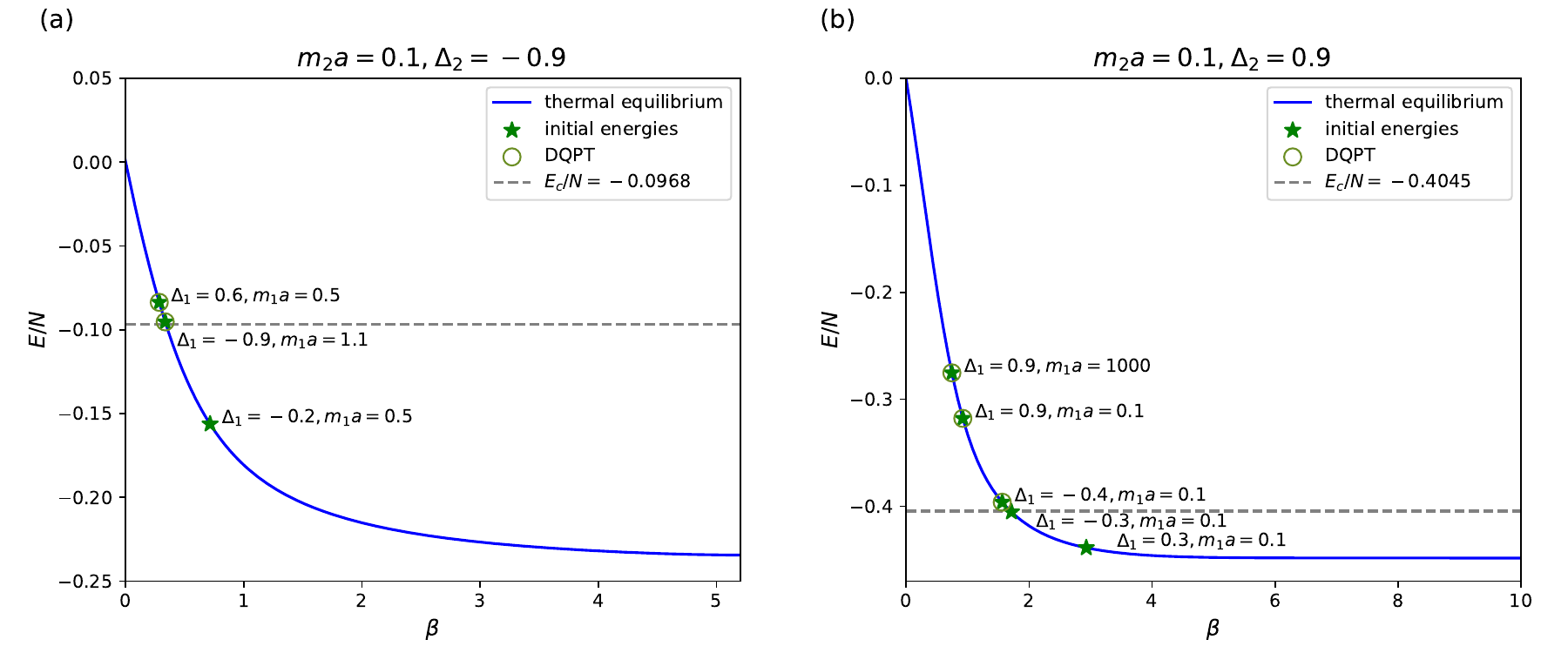}
\caption{Relation $E(\beta)$ between the mean energy and the inverse
temperature in the Gibbs state restricted to the zero magnetization sector for
$(m_2a,\Delta_2)=(0.1,-0.9)$ (left:critical phase) 
and $(0.1,0.9)$ (right:gapped phase). We display the
energy densities of several explored quenches using star symbols, and circles
indicate the initial states where our simulations reveal the occurrence of at
least one DQPT. Additionally, the dashed line denotes the lowest value of
energy density ($E_c/N$) at which a DQPT was observed (see also
Fig.~\ref{fig:energy_3D_plane}). For $(m_2a,\Delta_2)=(0.1,-0.9)$ (left),
$E_c/N=-0.0968$ and the corresponding inverse temperature 
$\beta_c\approx 0.34$; for $(m_2a,\Delta_2)=(0.1,0.9)$ (right), $E_c/N=-0.4045$
and $\beta_c\approx 1.70$.}
\label{fig:energyVsBeta}
\end{figure*}
As discussed in the previous section, the appearance of a DQPT after a certain
quench is not determined by the fact that the quench crossed an equilibrium
phase boundary. Hence the ground state of $H_2$ is not the most adequate
equilibrium reference for all initial states. Since the mean energy of the
initial state $E_1:=\bra{\Psi_{1}}|H_2 \ket{\Psi_{1}}$ is  typically much
higher than the ground-state energy of the quench Hamiltonian $H_2$, one does
not necessarily expect the time-evolved state to be close to the ground state
of $H_2$, even at the level of local observables. Instead, it is more 
meaningful to compare the local properties to those of the thermal equilibrium
state at the same energy density. If the system equilibrates (in the sense of
locally resembling an equilibrium ensemble), then the thermal state should
correspond to the maximal entropy state compatible with the conserved
quantities~\footnote{A detailed investigation of the equilibration 
process is beyond the scope of this work.}. Our particular model conserves 
energy and total magnetization, and
we are considering initial states in which the latter vanishes. Thus, the
relevant equilibrium state is the Gibbs state with the same energy $E_1$,
restricted to the subspace of vanishing total magnetization; namely (and up to
normalization) $\rho(\beta) \propto P_{s_z=0} \ e^{-\beta H_2} P_{s_z=0}$,
where $P_{s}=\sum_{n, {S}_z\ket{n}=s\ket{n}} \ket{n}\bra{n}|$ is the projector
onto the sector of total magnetization $s$. The inverse temperature
$\beta_1:=\beta(E_1)$ corresponding to the state is fixed by the condition
\beq
\label{eq:E_beta}
E_1=E(\beta_1):=\frac{\mathrm{tr} \left (
\rho_{\beta_1} H_2 \right )}{\mathrm{tr} \left ( \rho_{\beta_1}
\right )}.
\eeq

The determination of $\beta_1$ can be achieved using standard TNS methods,
exploiting the fact that the thermal equilibrium state for a local Hamiltonian
can be approximated efficiently by a matrix product operator
(MPO)~\cite{Hastings2006a,Molnar2015}. More concretely, given a value of the
inverse temperature $\beta$, the canonical purification of the thermal
equilibrium ensemble restricted to the vanishing magnetization sector can be
written as $\ket{\Psi(\beta)}:=
e^{-\beta H_2/2}\sum_{n, S_z\ket{n}=0} \ket{n}_{\mathrm{S}}\ket{n}_{\mathrm{A}}$,
where the sum runs over a basis restricted to the sector $S_z=0$, and the
subscripts $\mathrm{S}$ and $\mathrm{A}$ refer respectively to the system and
ancillary degrees of freedom. This state, also known as thermofield double
state, can be approximated by a MPS using standard tensor network
algorithms~\cite{Verstraete2008,Schollwoeck2011}. This is achieved by preparing
a maximally-mixed state of the system and the ancilla, expressed as a simple 
MPS, where each system site is entangled only with one, next neighbor, ancillary
site, and evolving it in imaginary time with an approximation of the operator
$e^{-\beta H_2/2}$. In order to restrict the ensemble to a sector of fixed
magnetization, we write the projector onto the sector $S_z=0$ as a MPO and use
as initial mixed state, instead of the identity. Additionally, this MPO can
also be applied to the intermediate state during the evolution to ensure that
truncation errors due to the approximations do not induce undesired
magnetization components. For a finite system of size $N$, the projector onto
vanishing magnetization has an exact MPO representation, with bond dimension
$N/2$. Applying the aforementioned TNS evolution algorithm in imaginary time to
the vectorized projector provides a MPS approximation to $\ket{\Psi(\beta)}$
for varying $\beta$, from which the thermal expectation value of a physical
observable $O$ can be computed directly as
\beq
\langle O \rangle_{\beta}:=
\frac{\mathrm{tr} (O \rho_{\beta})}{\mathrm{tr} \rho_{\beta}}=
\frac{\bra{\Psi(\beta)} |O \otimes \Id \ket{\Psi(\beta)}}
{\bra{\Psi(\beta)}\ket{\Psi(\beta)} },
\eeq 
where the second factor in the tensor product is the identity acting on the
ancillary system. In particular, computing the energy along the imaginary time
evolution provides an estimate of the function $E(\beta)$, from which we can
extract the value $\beta_1$ corresponding to a given initial state by inverting
it numerically.

For concreteness, in the rest of the section we focus on two quenches,
corresponding to Hamiltonians in each of the phases (see
Fig.~\ref{fig:equ_phase_dqpt}). Here we choose $(m_2 a,\Delta_2)=(0.1,-0.9)$
in the gapless phase and $(m_2 a,\Delta_2)=(0.1,0.9)$ in the gapped one.
For each of them we approximate the thermal equilibrium state in a system of
$N=300$ sites, sufficiently large to neglect finite-size effects in this
analysis. The corresponding $E(\beta)$ curves are shown in
Fig.~\ref{fig:energyVsBeta}.

\begin{figure*}[t]
\centering
\includegraphics[width=\linewidth]{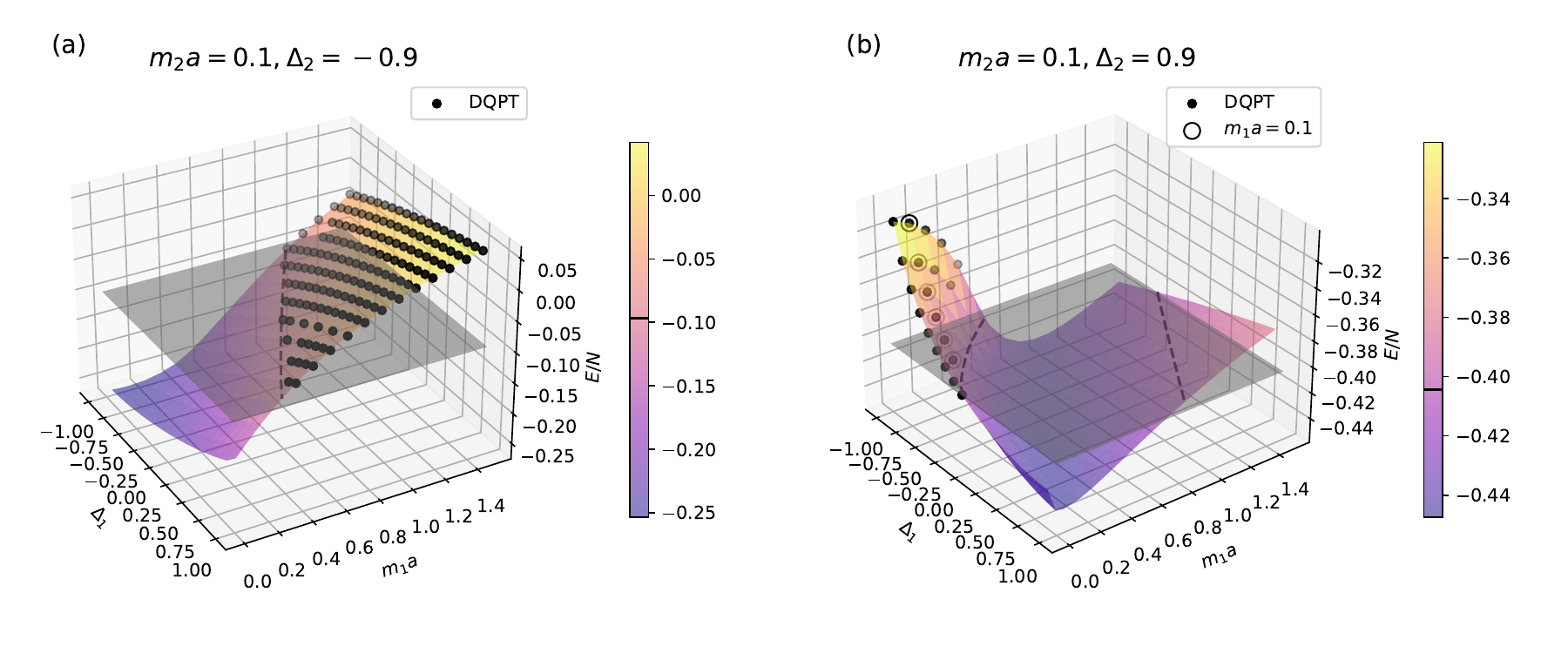}
\caption{ Energy densities  corresponding to ground states of varying $m_1 a$
and $\Delta_1$, as a function of these parameters, for the quench to 
the critical phase $(m_2a=0.1,\,\Delta_2=-0.9)$ (left) and the 
gapped phase $(m_2a=0.1,\,\Delta_2=0.9)$ (right).
In each case, the dots indicate where DQPT has been observed. The horizontal
plane shows the energy threshold, $E_c/N$, which is also marked as a black line
on the color bars. On the right panel, the circles indicate the family of
quenches from constant $m_1 a=0.1$.}
\label{fig:energy_3D_plane}
\end{figure*}
For each of the simulated initial states, we can now extract the effective
temperature from these curves. The ground states of the different initial
Hamiltonians $H_1$ correspond to initial states with different effective
temperatures. Since the latter only depends on the energy $E_1$, states in very
distant regions of the equilibrium phase diagram may have the same temperature.
As an illustration, we show the energy densities of some of the explored
quenches with star symbols in Fig.~\ref{fig:energyVsBeta}. We furthermore
indicate with circles the initial states for which our simulations show the
appearance of (at least one) DQPT. From these results we observe that for DQPTs
to be present, the effective temperature of the initial state needs to surpass
a certain threshold. This seems to suggest that a decisive factor for the
occurrence or not of a DQPT is the amount of excitation energy in the initial
state. 

To verify this observation, we estimated the value of $E_1$  for a grid of
$(m_1a,\Delta_1)$ values, $0< m_1 a\leq 1.5$ and $-1\leq \Delta_1\leq 1$, and
identified the regions of parameters for which the threshold (denoted as $E_c$)
is surpassed (see Fig.~\ref{fig:energy_3D_plane}). We conducted this study for
both quench Hamiltonians mentioned above. For the gapless Hamiltonian,
$(m_2 a,\Delta_2)=(0.1,-0.9)$, the value of $E_1$ increases smoothly with both
$m_1 a$ and $\Delta_1$, and the plane of constant {$E_c/N\approx-0.0968$},
corresponding to the lowest value for which a DQPT was observed.  As indicated
in Fig.~\ref{fig:energy_3D_plane}(a), this divides the parameter space in two
connected regions. Consistent with this, we find DQPTs for all initial states
sampled from the region with higher energy.

In contrast, for the quench into the gapped phase, $(m_2 a,\Delta_2)=(0.1,0.9)$,
the dependence of $E_1$ on $(m_1 a,\Delta_1)$ is not monotonic. We find two
disjoint regions of parameters with energies above the identified threshold
$E_c/N\approx-0.4045$, lying on opposite corners of the parameter space as can
be seen in Fig.~\ref{fig:energy_3D_plane}(b). One region with negative $\Delta_1$
and small $m_1 a$ (overlapping but not coinciding with the critical phase, as
shown in Fig.~\ref{fig:equ_phase_dqpt}),  and the other region with positive
large $\Delta_1$ and $m_1 a$, deep within the gapped regime. We simulate
quenches from both regions, as well as intermediate parameters with energy $E_1$
below the threshold. For the first high-energy region, i.e., negative $\Delta_1$
and small $m_1 a$, we consistently observe DQPTs for energies above the
threshold. The second region, with large positive $\Delta_1$ and $m_1 a$,
however, seems to exhibit DQPTs only at much later times than the first. An
extreme case is shown in Fig.~\ref{fig:selected_quenches}(d), for the quench from
$m_1a=1000$, $\Delta_1=0.9$, for which we observe a DQPT only close to $t\approx 10$
. We observe that as $m_1 a$ values decrease, the
time at which the first DQPT is observed increases even more, so that we need to run 
long-time simulations to analyze this region. Due to limitations imposed
by truncation errors, we can reliably explore times up to only $t\lesssim 14$.
Within this window, we only observe DQPTs for points in the parameter regime
with masses outside our original grid. We presume that longer times would
identify DQPTs also in the rest of the region above the threshold.

\begin{figure*}[ht]
  \centering
  \includegraphics[width=\linewidth,trim = 20 0 0 0]{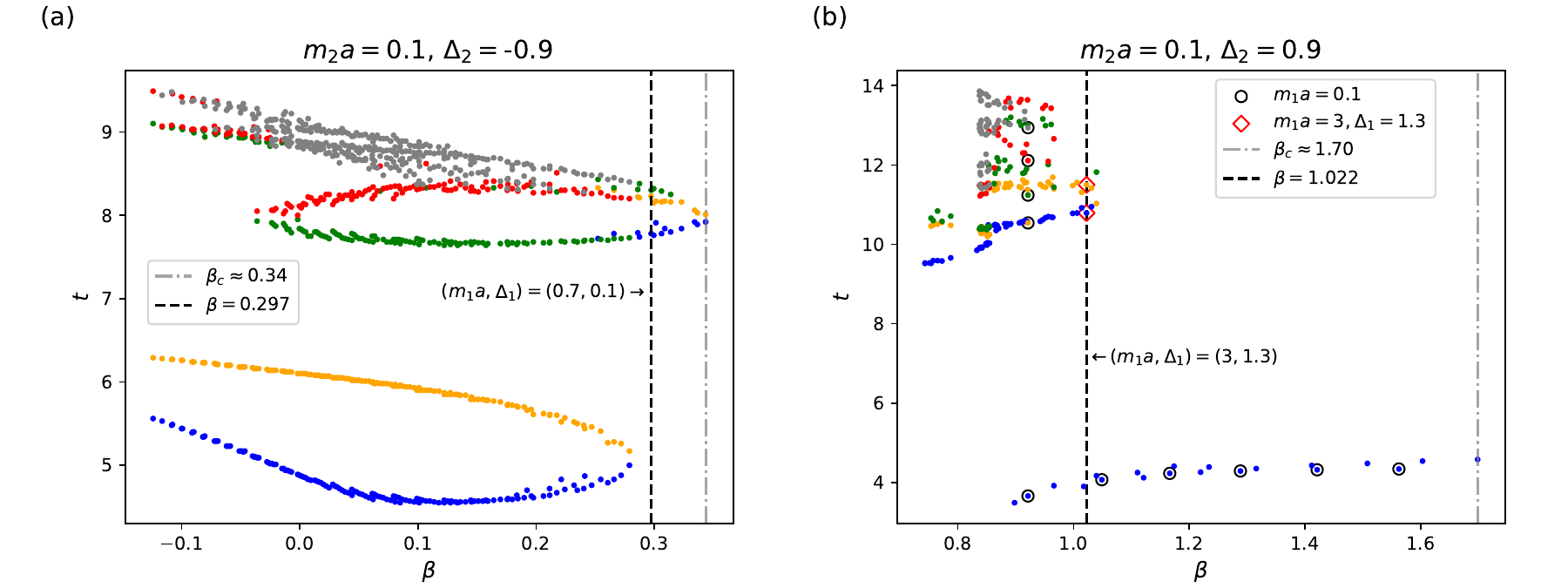}
\caption{
The DQPTs on the inverse temperature--time plane. The first, second, third, and
fourth occurrences of DQPTs are colored blue, orange, green, and red,
respectively. Gray indicates occurrences beyond the fourth. The gray
dash-dotted line denotes the critical inverse temperature $\beta_c$.
(a) We simulate the evolution parameters $m_2 a=0.1$ and $\Delta_2=-0.9$ (critical phase) and
perform a scan for the initial parameters
($m_1 a=0.1, 0.2,\dots ,1.5$; $\Delta_1=-0.9, -0.8,\dots,0.9$).
For example, the dashed line is the case for $m_1a=0.7$ and $\Delta_1=0.1$,
which is also specifically shown in Fig.~\ref{fig:selected_quenches}(b).
(b) The simulation results for the evolution parameters $m_2 a=0.1$ and
$\Delta_2=0.9$ (gapped phase), scanning for the initial parameters as follows:
($m_1 a=0$; $\Delta_1=-0.9, -0.8,\dots, -0.1, 0.9$), 
($m_1 a=0.1, 0.2,\dots ,1.5$; $\Delta_1=-0.9, -0.8,\dots,0.9$), 
($m_1 a=1, 2,\dots ,9$; $\Delta_1=1,1.1,1.3,1.5,\dots,1.9$), and 
($m_1 a=10,20,30,40,60,80,100,1000$; $\Delta_1=0.9$). 
These large masses are chosen to explore the large $\Delta_1$, large $m_1 a$ corner 
of Fig.~\ref{fig:energy_3D_plane}(b), while ensuring that DQPT occurs within our 
simulation window ($t<14$). 
The black dashed line and
the red diamond symbols represent the case for $m_1a=1.3$ and $\Delta_1=3$,
where DQPT does not occur at the lowest branch near $t\approx 4$. 
}
\label{fig:beta_t_branch}
\end{figure*}
This exploration of the parameter space indicates that for states with similar
effective temperature, but corresponding to very different values of
$(m_1a,\,\Delta_1)$, the DQPTs may occur in  very different ways. In order to
obtain a more complete picture, we plot in Fig.~\ref{fig:beta_t_branch}, the
times at which we observe DQPTs as a function of the effective temperature of
the initial state for all the simulated cases. For each of the initial states,
we determine $\beta$ from the energy expectation value
$\bra \Psi_{1} | H_2 \ket {\Psi_{1}}$ through Eq.~\eqref{eq:E_beta}. The figure
shows not only the first observed nonanalyticity (blue dots), but the times of
all subsequently observed DQPTs (with colors indicating the number of the DQPT).
These figures are reminiscent of the nonanalyticities shown by the complex
Loschmidt rate, when defined on the complex $t$ plane for a fixed initial
state~\cite{Heyl2018review}. In such case, the nonanalyticities of the rate in
the complex-time plane are analogous to Fisher zeros of a {boundary} partition
function and determine the presence and time of occurrence of
DQPTs~\cite{Peotta2021cumulant,McCulloch2023dqpt}. In our case, the initial
state varies and we do not evolve it in imaginary time, but the results suggest
that our effective temperature does play an equivalent role, allowing us to
explore such complex plane from purely real-time data. 
However, we also observe that some initial states are not sensitive to some of
the DQPT \emph{branches}. For instance, if we select $(m_1a, \Delta_1)=(3, 1.3)$
and quenches to $(m_2a, \Delta_2)=(0.1, 0.9)$ [indicated by the black dashed
line and red diamond marks in Fig.~\ref{fig:beta_t_branch}(b)], then the DQPT does not
occur at the lowest branch near $t\approx 4$. The generality of this connection
for other models thus needs further investigation.

\subsection{String correlator and DQPT}

Some DQPTs have been connected to equilibrium phase diagrams~\cite{Heyl2012dqpt,
Heyl2018review, Karrasch_2017,Hashizume2022}. In Ref.~\cite{Banuls:2019hzc}, we
identified the ground-state phase structure by analyzing the behavior of the
string correlator \eqref{eq:Cstring} (see Sec.~\ref{sec:equil_phase_struct}).
Namely, the zero-temperature gapless phase is characterized by the string
correlator decaying to zero at large distances ($C=0$), while the gapped phase
evinces a decay of this correlator to a nonzero value ($C>0$). In the present
work, we use similar methodology to connect the occurrence of DQPT to the
thermal string correlator, pertaining to the effective temperature of the
initial state. As in Ref.~\cite{Banuls:2019hzc}, we fit the distance-dependent
string correlators with the power-exponential ansatz given by
Eq.~\eqref{eq:corr_ansatz}. To establish the connection to DQPT, we note that,
keeping the mass fixed, $m_1a=m_2a=0.1$, and varying $\Delta_1$, each value of
the coupling corresponds to a different effective $\beta$, as illustrated in
Fig.~\ref{fig:energyVsBeta}(a) (for $\Delta_2=-0.9$) and 
Fig.~\ref{fig:energyVsBeta}(b) (for $\Delta_2=0.9$).

\begin{figure}[ht]
  \centering
    \includegraphics[width=1\columnwidth]{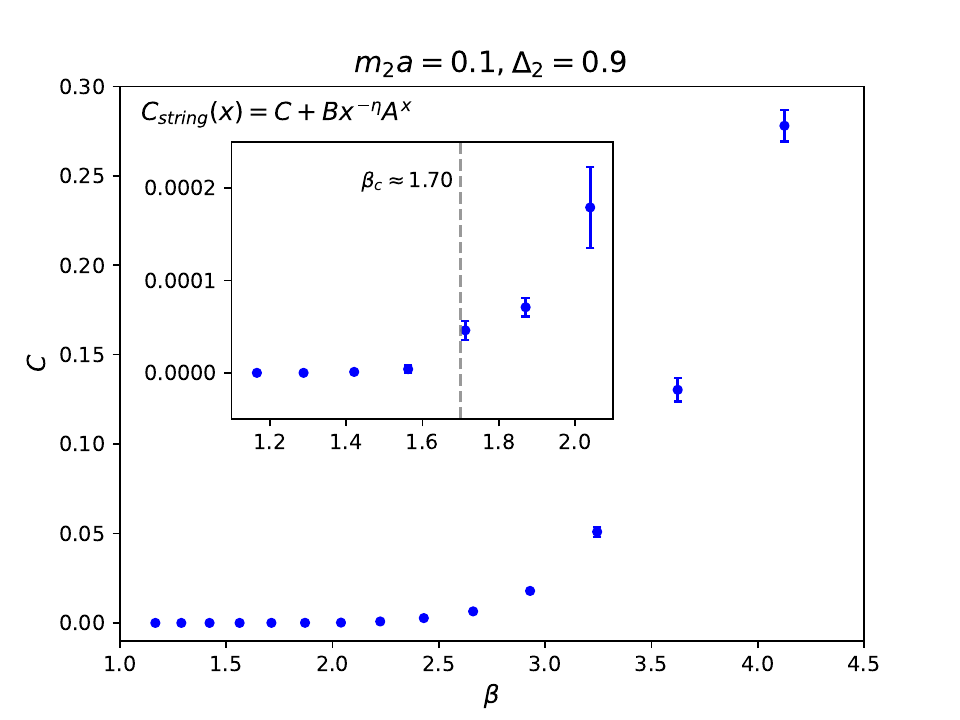}
    \caption{ 
    Value of the constant $C$, defined in Eq.~\eqref{eq:corr_ansatz}, that 
    characterizes the long-distance behavior of the string
    correlator~\eqref{eq:Cstring} for the thermal equilibrium state of the
    Hamiltonian $(m_2a,\Delta_2)=(0.1,0.9)$ as a function of the inverse
    temperature $\beta$. The gray dashed line in the inset denotes the critical
    inverse temperature $\beta_c$.
    }
    \label{fig:finiteT_phase_diagram}
\end{figure}

We start the discussion with the latter, $\Delta_2=0.9$. We observe the
following correspondence (see also Fig.\ \ref{fig:finiteT_phase_diagram}):\\
\hspace*{15 pt}(i) \quad for $\beta\leq1.563$ ($\Delta_1\leq-0.4$), the 
constant $C$ is consistent with zero [at $\beta=1.563$, $C=0.000004(4)$] 
and DQPT is observed,\\
\hspace*{15 pt}(ii) \quad  for $\beta\geq1.712$ ($\Delta_1\geq-0.3)$, a 
nonzero constant emerges [at $\beta=1.712$, $C=0.000046(10)$] and we find no DQPT.

The observed correspondence pertains to DQPTs pictured in the lowest branch of
Fig.~\ref{fig:beta_t_branch}(b), explicitly marked with circles when
$m_1a=m_2a$. Thus, this kind of DQPT is related to a substantial change in the
properties of the thermal state, characterized by a different value of $C$. The 
former/latter may be the thermal counterpart of the zero-temperature
critical/gapped phase, but a detailed characterization of finite-temperature
phases of the model is beyond the scope of our work.

The picture is somewhat different for $\Delta_2=-0.9$.
In this case, $C=0$ at all temperatures, regardless of the occurrence of DQPT,
found at {$\beta\lesssim0.34$} [see Fig.~\ref{fig:beta_t_branch}(a)]. Note,
however, that none of the DQPTs represented in Fig.~\ref{fig:beta_t_branch}(a)
occur at $m_1a=m_2a$, i.e., a prerequisite for DQPT at $\Delta_2=-0.9$ is to
have a quench in fermion mass.  
Thus, for the final Hamiltonian in the critical phase 
$(m_2a,\,\Delta_2)=(0.1,\,-0.9)$, the analyzed string correlator does not exhibit 
a similar correlation to the presence of DQPTs as the one observed for the 
quench into the gapped Hamiltonian.
This might be understood because the asymptotic value of the string correlator 
detects the order present in the ground state of the gapped phase, which 
is destroyed as the temperature (energy) increases, whereas the ground state 
in the critical case does not exhibit such order.
On the other hand, the DQPT threshold detected by the correlator in the gapped 
quench corresponds to the abrupt edge of the lower DQPT branch in 
Fig.~\ref{fig:beta_t_branch}(b), a feature that does not appear either in 
the corresponding Fig.~\ref{fig:beta_t_branch}(a) for the critical quench. 
Although we conclude that the thermal string correlators defined in 
Eq.~\eqref{eq:Cstring} are not universal indicators for the possibility of 
DQPTs, a more detailed investigation is needed to fully characterize which 
DQPTs are associated to changes in these quantities and to determine whether 
other thermal observables can detect the different boundaries of DQPT branches 
observed in Fig.~\ref{fig:beta_t_branch}. We thus leave the investigation of
finite-temperature phases and their potential relation to DQPTs for future work.
\section{Conclusion}
\label{sec:conclusion}
Using standard TNS methods, we have investigated the occurrence of DQPTs in the
lattice discretization of the Thirring model, which corresponds to a spin-$1/2$
XXZ chain coupled to a uniform and a staggered magnetic field, with two free
dimensionless parameters, respectively related to the coupling and the mass.
Using the uniform MPS ansatz to work directly in the thermodynamic limit, we
have simulated quenches where the evolution is given by a constant Hamiltonian
and the initial state is chosen as the ground state for different parameters,
in the same or a different phase. We then studied the time evolution of the
spectrum of the mixed transfer operator, whose dominant eigenvector directly
determines the Loschmidt rate.  DQPTs are found when another eigenvalue crosses
with the largest one.
 
We have identified a threshold in the energy density of the initial state below
which DQPTs do not occur, at least during the finite times reachable by our
simulations. The threshold is observed both for quenches in the gapless and the
gapped phases. The energy density of each initial state corresponds to an
effective temperature: that of the thermal equilibrium state with the same
energy per spin, restricted to the sector of zero magnetization. This thermal
state is the one expected to describe the equilibrium values of local
observables after the quench in the long-time limit, since energy and
magnetization are the only local conserved quantities in the problem.

We have analyzed the position of each observed DQPT (or nonanalyticity of the
Loschmidt rate) in the time-effective-temperature plane. This reveals
structures of nonanalyticities reminiscent of the  zeros in the complex-time
plane of the Loschmidt rate defined for a fixed initial state. Our results show
that DQPTs for different initial states lie in multiple well-defined branches.
Initial states in different phases can appear on the same branch, which thus is
not related to the zero-temperature phase diagram. All DQPT branches start at
some minimal effective temperature. In most cases, the lowest temperature edge
of a branch corresponds to a smooth merge with another one. However, in the
case of the quench with the gapped Hamiltonian, we identify a particular
branch, with DQPTs happening at the shortest times, with a sharp end at
$\beta_c\sim 1.7$.
We have found that this energy threshold corresponds very accurately with a
substantial change in the properties of the thermal equilibrium states at the
corresponding effective temperature. More concretely, below this temperature, a
certain string correlator attains a nonzero value at long distances, a
property that also characterizes the ground state in the gapped phase. For this
branch, we have thus established a relation between the presence of DQPTs and
the equilibrium phase diagram at finite $T$. We have not found a similar
correspondence for the smooth thresholds of branches that merge into each other.

Our study thus suggests that observing the dynamics and analyzing the DQPTs
from different initial states can probe properties of the finite-temperature
phase diagram, in particular when a sharp threshold is found for the occurrence
of these nonanalyticities. Since our simulations are limited to finite-time
windows, due to the accumulation of truncation error, and we have not mapped
out the complete thermal equilibrium properties at all temperatures, further
investigation is needed to determine whether the other observed branches
correspond also to some quantifiable equilibrium property. It is also
interesting to investigate whether similar effects appear in other spin models,
and the precise correspondence between Fisher zeros and the DQPT lines in our
time-effective temperature representation.

\acknowledgments
The authors thank I. McCulloch, L. Tagliacozzo, M. Fagotti, L. Barbiero, and 
N. Cuzzuol for very useful discussions. This work was partly supported by the 
DFG (German Research Foundation) under Germany's Excellence Strategy 
-- EXC-2111 -- 390814868 and Research Unit FOR 5522 (Grant No. 499180199) and 
by the EU-QUANTERA project TNiSQ (BA 6059/1-1), as well as Taiwanese NSTC 
Grants No. 110-2112-M-002-034-MY3, No. 112-2112-M-A49-021-MY3, 
No. 112-2119-M-007-008, and No. 113-2119-M-007-013. Numerical computations 
were performed on HPC facilities at National Taiwan University and National 
Yang Ming Chiao Tung University.

\section*{Data availability}
The data and code to generate the figures are available in Github 
\cite{Thirring_data}.

\appendix

\section{Penalty Term and Matrix Product Operator}
\label{appdx:penalty_mpo}
Since the $z$ component of the total spin corresponds to the total fermion
number in the Thirring model, and we are only interested in the zero-charge
sector. To enforce this, we introduce a penalty term in the Hamiltonian,
\beq
\label{eq:H_penalty}
 \bar{H}^{\text{penalty}}=H+ \lambda \left( \sum_{n=0}^{N-1} S_{n}^{z} \right)^2.
\eeq
Here $\lambda$ should be sufficiently large (100 in this work) to ensure that
the ground state obtained {\it via} a variational search is in the sector of
vanishing total $S^{z}$~\cite{Banuls:2013jaa}, that is,
$\la{S_{tot}^{z}}\ra=\la{\sum_n{S_{n}^{z}}}\ra=0$. This enables us to interpret
our results in terms of the dual SG theory and the XY model. However, the
penalty term will not be turned on during the real-time evolution simulations
since $\la S_{tot}^z\ra$ is a conserved quantity. In our simulations, this
quantity is well preserved, i.e., $\la S_{tot}^z(t)\ra=0$.

In TN algorithms, it is useful to express the Hamiltonian as the MPO. 
The Hamiltonian of Thirring model in Eq.~\eqref{eq:H_penalty}
can be expressed as an MPO:
\beq
 \label{eq:MPO}
\begin{split} 
&W^{[n]}  =
\begin{pmatrix}
\mathbb{I} & -\frac{1}{2}S^{+} & -\frac{1}{2}S^{-}  & 2\lambda S^z & \Delta S^z 
& \beta_n S^z+\alpha \mathbb{I}\\
0 & 0 & 0 & 0 & 0 & S^{-} \\
0 & 0 & 0 & 0 & 0 & S^{+} \\
0 & 0 & 0 & \mathbb{I} & 0 & S^{z} \\
0 & 0 & 0 & 0 & 0 & S^{z} \\
0 & 0 & 0 & 0 & 0 & \mathbb{I} \\
\end{pmatrix}
\end{split}.
\eeq
where
$\beta_n = \Delta + (-1)^n\tilde{m_0}a,\,
\alpha = \frac{1}{4}\left(\lambda + \Delta \right)$.
The $\beta_n$ contains the staggered term. To simplify the
simulation, we also combine two-site MPOs into one site, as we did with the
uMPS, represented as $W_1 W_2\rightarrow W$.

\section{Brief Introduction of VUMPS and TDVP}

\begin{figure}[ht]
  \centering
  \includegraphics[width=1.\linewidth]{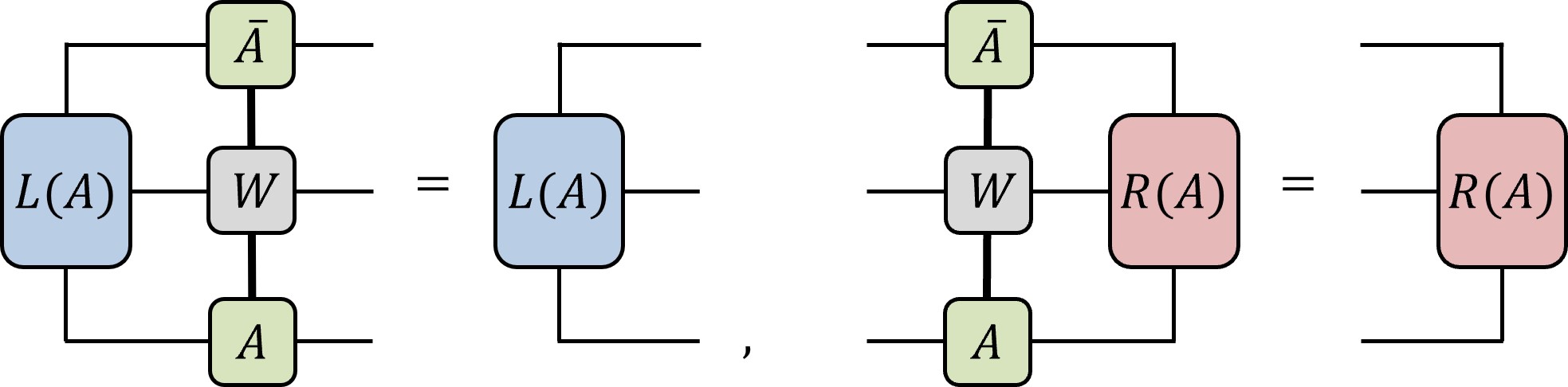}
  \caption{The fixed-point equations of the boundary tensors $L$ and $R$.}
  \label{fig:fixed_pt}
\end{figure}

To prepare the ground state in uMPS form, we use the VUMPS
algorithm~\cite{Zauner2018vumps}. In VUMPS, we need to construct the effective
Hamiltonian through the MPO and the boundary tensors $L, R$, which
satisfy the fixed-point equations (see Fig.~\ref{fig:fixed_pt}). 

The tensor diagram of the effective Hamiltonian operating on the unit cell can
be expressed as:
\begin{equation}
\label{eq:H_eff}
 \includegraphics[scale = 0.7]{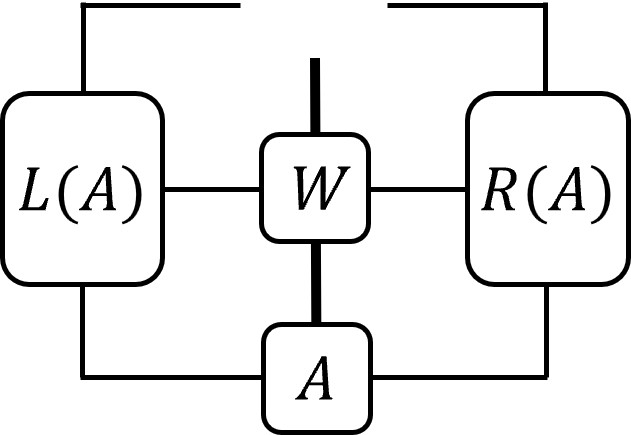} \, .
\end{equation}
The boundary tensors $L$ and $R$ are functions of $A$. We could use the Lanczos
algorithm to obtain the optimal $A$ and then update $L$ and $R$ until it reach
the ground state. In our simulation, we turned on the penalty term $\lambda$ in
Eq.~\eqref{eq:MPO} to guarantee that the ground state we obtained is in the
sector of vanishing total $\la S_z\ra$.

\begin{figure}[ht]
  \centering
  \includegraphics[width=.6\linewidth]{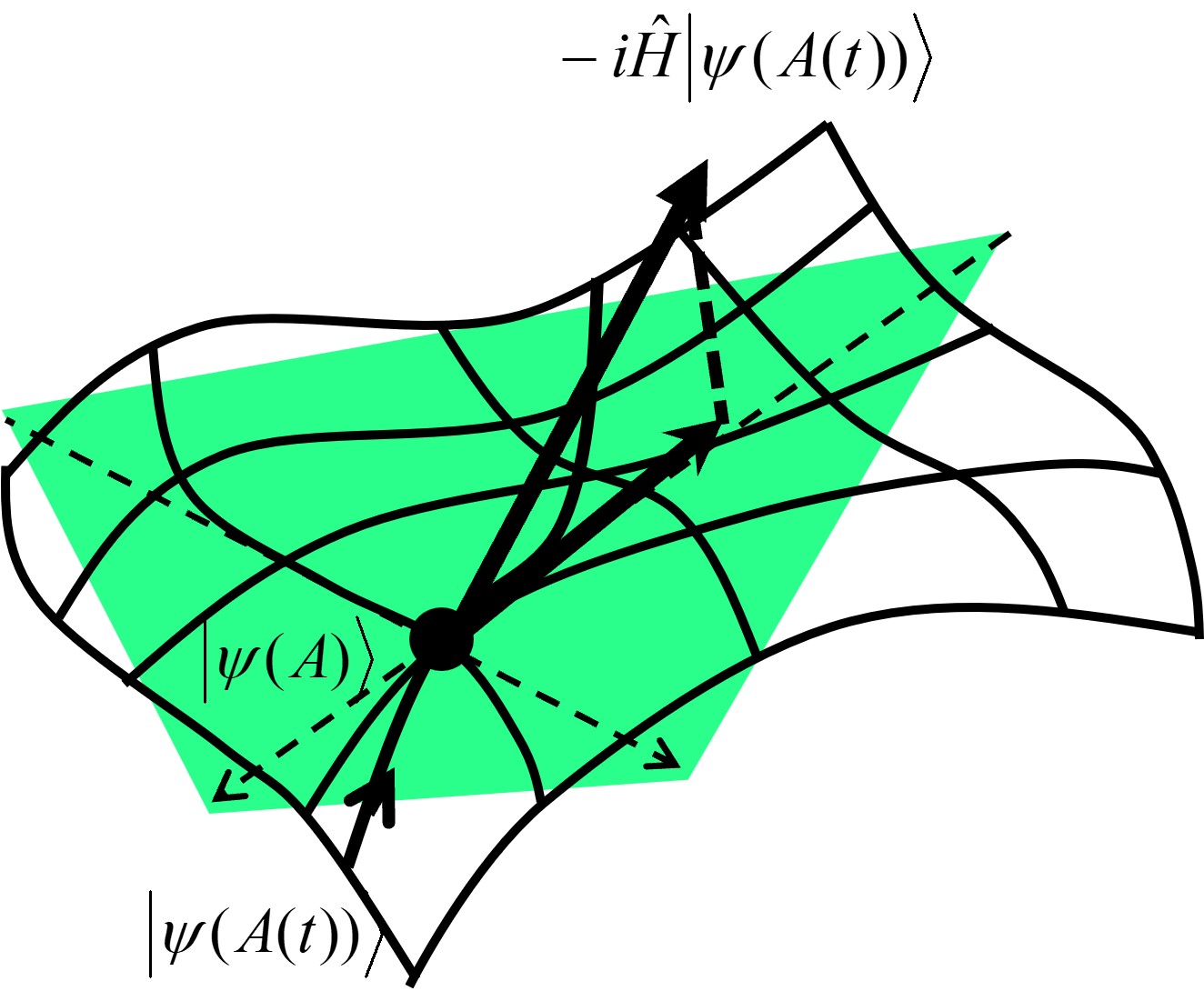}
  \caption{An illustration of the TDVP. The curve is the manifold of the uMPS
  and the colored plane is the tangent space of the current state.}
  \label{fig:tdvp_concept}
\end{figure}

For real-time evolution, we choose the TDVP algorithm, which preserves
conserved quantities well \cite{Haegeman2011itdvp}. TDVP corresponds to a
projection of the evolution state of the Schr$\ddot{\text{o}}$dinger equation
onto the tangent space of the uMPS manifold $\mathcal{M}$ at the current state
(see Fig.~\ref{fig:tdvp_concept}). We can solve the time derivative of the
uMPS $|\Psi(A)\ra$ through the TDVP equation:
\beq \label{eq:tdvp}
\frac{\partial}{\partial t}|\Psi(A)\ra=
-i\hat{P}_{T_{|\Psi(A)\ra}\mathcal{M}}\hat{H}|\Psi(A)\ra.
\eeq
The left-hand side of Eq.~\eqref{eq:tdvp} gives the time derivative of the
tensor $A$, while the right-hand side gives the projection result:$-iB$, where
$B$ is also a tensor that has the same shape as tensor $A$. Thus, we will get
$\dot{A} = -iB$, which can be solved using the Runge-Kutta (RK4) method. Note
that for real-time evolution, we turn off the penalty term $\lambda$ in
Eq.~\eqref{eq:MPO} (setting $\lambda=0$).

%

\addcontentsline{toc}{chapter}{Bibliography} 
\bibliography{refs} 
 
\end{document}